\documentclass[preprint,12pt]{elsarticle}

\usepackage{graphics}
\usepackage{graphicx}
\usepackage{epsfig}

\usepackage{amssymb}
\usepackage{amsthm}

\usepackage{txfonts}  
\usepackage{natbib}
\usepackage{times}

\newcommand {\arcsec}{\mbox{$^{\prime\prime}$}}

\newcommand{\msun}{{\,\rm M}_{\odot}}

\journal{New Astronomy Reviews}

\begin{document}

\begin{frontmatter}




\title{Exoplanet search with astrometry}


\author{R. Launhardt}

\address{Max Planck Institute for Astronomy, K\"onigstuhl 17, 69117 Heidelberg, Germany}

\begin{abstract}

Searching for extrasolar planets by direct detection  
is extremely challenging for current instrumentation. 
Indirect methods, that measure the effect of a planet on its host star, 
are much more promising and have indeed led to the discovery of nearly all 
extrasolar systems known today. While the most successful method thus far is 
the radial velocity technique, new interferometric instruments like PRIMA at the VLTI 
will enable us to carry out astrometric measurements accurate enough to 
detect extrasolar planets and to determine all orbital parameters, 
including their orbit inclination and true mass.
In this article I describe the narrow-angle astrometry technique, 
how it will be realized with PRIMA, what kind of planets we can find, 
and what kind of preparatory observations are required.

\end{abstract}


\end{frontmatter}



\section{Introduction} \label{sec-intro}

The discovery of planets orbiting stars other than our Sun is one of the greatest 
scientific and philosophical achievements of our time. Aside from providing us with 
a wealth of information to understand the formation and structure of planetary systems 
in a universal context, it captures the interest of both scientists and the public 
with the prospect of finding life in the Universe. Triggered by the discovery of the 
first planet orbiting another solar-like star in 1995 \cite{mayor1995}, 
more than 300 extrasolar planets have been discovered, the majority of 
them by using the radial velocity method. The dominant role of this technique is however 
being eroded by the arrival of new facilities. 

The European Southern Observatory (ESO) is currently integrating and testing 
PRIMA, the instrument for {\bf P}hase-{\bf R}eferenced {\bf I}maging and 
{\bf M}icro-arcsecond {\bf A}strometry 
\cite{delpl2000,delpl2006,delpl2008,derie2003} at the
Very Large Telescope Interferometer on Cerro Paranal in Chile. 
The VLTI consists of four stationary 8.2-m VLT ''Unit Telescopes'' (UTs), four movable 
1.8-m ''Auxiliary Telescopes'' (ATs), and six long-stroke delay lines (DLs). 
It provides baselines of up to 200\,m and covers a wavelength range 
that extends from the near infrared (1\,$\mu$m) to 13\,$\mu$m 
\cite{schoel2007,schoel2009}. 

PRIMA will implement the dual-feed capability at the VLTI, at first for two UTs or ATs, 
to enable simultaneous interferometric observations of two objects that are separated 
by up to 1 arcmin.
PRIMA consists of
\begin{enumerate}
\item 
Star Separators (STS) for each telescope (at first two) that separate two sub-fields 
(with a star in each) from the telescope field-of-view 
and send them as collimated beams through the interferometer, 
\item 
Differential Delay Lines (DDLs) that compensate for the optical path difference (OPD) 
between the two stars,
\item
Fringe Sensor Units (FSUs), which combine the beams and detect the interferometric fringes, and 
\item
an end-to-end laser metrology system (PRIMET) that measures the OPDs within the interferometer.
\end{enumerate}
PRIMA is designed to perform narrow-angle astrometry 
in K-band with two ATs as well as external fringe-tracking for phase-referenced 
aperture synthesis imaging with UTs or ATs and instruments like AMBER \cite{petrov2000} 
and MIDI \cite{lein2003}. 

The PRIMA facility will soon provide us with the infrastructure to carry 
out an astrometric search for extrasolar planets. This will both complement some weaknesses 
inherent to the radial velocity method as well as open new discovery spaces. 
In order to speed up the full implementation of the astrometric capability and to carry 
out a large astrometric planet search program, a consortium lead by the Observatoire 
de Gen\`eve (Switzerland), Max Planck Institute for Astronomy, and Landessternwarte 
Heidelberg (Germany), has built Differential Delay Lines (DDLs) for PRIMA 
\cite{pepe2008} and is currently developing the astrometric observation preparation 
and data reduction software \cite{elias2008}. 
When PRIMA becomes fully operational in 2010, the consortium will to use it with two ATs to carry 
out a systematic astrometric Exoplanet Search with PRIMA (ESPRI) 
\cite{lau2005,lau2007,lau2008a,lau2008b,lau2008c,quir2004,reff2006}.


\section{Extrasolar planets - overview on search methods} \label{sec-planets}

There are various methods to detect planets around other stars \cite{perry2000}.
One can distinguish two detection principles: 
\begin{enumerate}
\item {\bf Direct methods} detect photons (or other signals) that come directly from the planet.
\item {\bf Indirect methods} measure the influence of a planet on its host star.
      The presence of a planet is deduced via models.
\end{enumerate}


\begin{figure}[hbt]
\begin{center}
\includegraphics[width=0.52\textwidth,angle=-90]{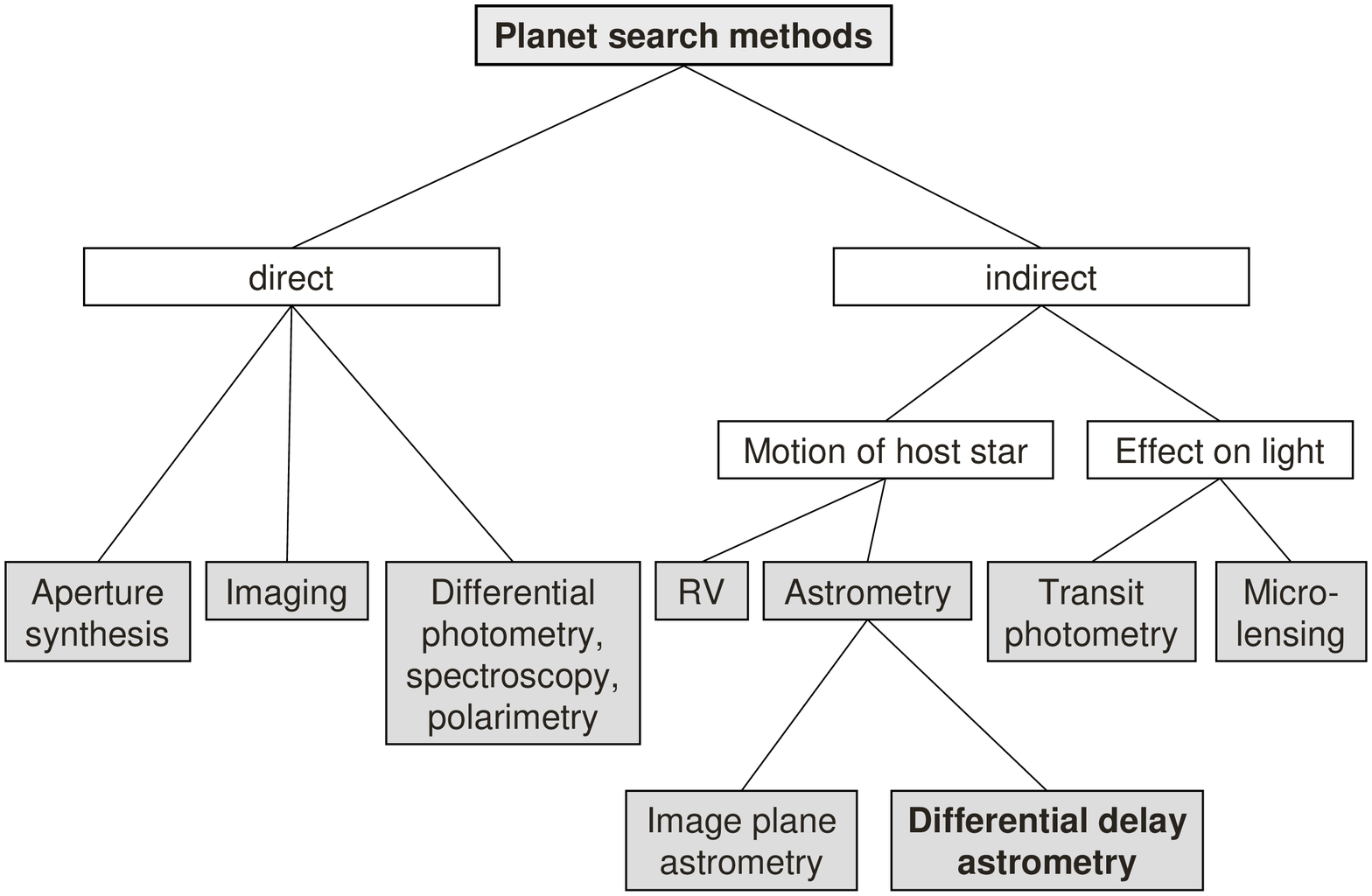}
\caption{\label{fig-methods}
Schematic overview of the most widely used planet search methods
}
\end{center}
\end{figure}


Direct imaging has to overcome the problems caused by the huge brightness contrast 
(between $10^6:1$\ and $10^9:1$, depending on planet, star, and wavelength) 
and the small angular separation between star and planet (1\,AU at 10\,pc distance corresponds to 
0.1\arcsec\ projected separation). 
Therefore, even the most advanced current-day and near-future telescopes and cameras 
can detect only exceptionally bright planets in large orbits around nearby stars with low mass and 
luminosity.
Other direct methods that include differential imaging and spectroscopy, possibly combined with 
polarimetry, are now sufficiently advanced to lead to detections of certain planetary systems 
\cite{kalas2008,marois2008}.
Aperture synthesis imaging with interferometers, that could include nulling of the star light, 
may eventually become a method to directly detect Earth-like planets. But, this technology 
is not yet advanced enough. 
 
At present, indirect methods are better-suited and more widely used to search 
for extrasolar planets. 
The two principle indirect methods measure 
{\it (i)} the effect of the planet on the star light, and 
{\it (ii)} the effect on the motion of the host star (see Fig.\,\ref{fig-methods}).
Methods that measure the effect on the light from the star star include:
\begin{itemize}
\item
{\bf Transit photometry}, which measures the dimming of the light when a planet 
  moves in front of its host star. 
  This method is sensitive only to planetary systems that 
  have their orbital planes aligned with the line of sight towards the Earth \cite{charb2007}.
\item
{\bf Microlensing}, which measures the apparent magnification of the light from a 
  background star due to relativistic light-bending when a planet moves in the line of sight 
  between a distant background star and the Earth \cite{bennet2002}.
\end{itemize}

\begin{figure}
\begin{center}
\includegraphics[width=0.5\textwidth,angle=0]{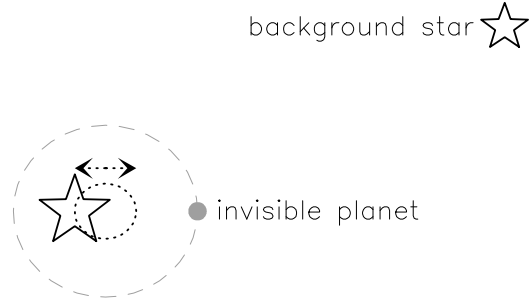}
\caption{\label{fig-astrom0}
Detecting a planet via astrometric measurements of the host star's position.
Both planet and star orbit around their barycenter, which is, due to 
the large mass ratio between the two, usually located very close to the star. While the planet 
itself remains invisible, the ``wobble'' of its host star in the plane of the sky 
can be measured directly. A background star serves as position reference.
}
\end{center}
\end{figure}

Since the gravitational influence of a planet on its host star is well-defined by 
Kepler's laws, planet searches today mostly measure the motion of a star as 
it orbits around the barycenter of the star\,--\,planet system. 
Due to the large mass ratio between star and planet ($\approx1,000:1$\ for Sun and Jupiter 
and $\approx330,000:1$\ for Sun and Earth), the common center of mass is 
usually located very close to the star (the center of mass between Sun and Jupiter is 
located approximately on the surface of the Sun). 
Consequently, the wobble of the star is quite small and not easy to measure.
Although we cannot directly measure the three-dimensional motion of a star on the sky, there 
are two alternative approaches of measuring different components of this motion:
\begin{itemize}
\item 
 The {\bf Radial Velocity (RV) technique} determines the velocity component of the star's motion 
 along the line of sight by measuring the Doppler shift of narrow spectral lines \cite{cumming1999}.
 It measures only one of the three space components of the star's motion and, 
 therefore, leaves the inclination of the orbit undetermined and provides only a lower limit 
 on the mass of the planet. Nevertheless, it is thus far the most successful technique that has lead 
 to the discovery of most of the more than 300 known extrasolar planets \cite{schnei1995}.
\item 
 The {\bf astrometric technique} measures the change of the projected position 
 of a star in the plane of the sky (see Fig.\,\ref{fig-astrom0}). 
 The amplitude of this positional ``wobble'' is very small and not easy to measure. 
 For example, Jupiter causes the Sun to move by $\approx1$\,mas over 12\,yrs when viewed from a 
 distance of 10\,pc. Since the astrometric method measures two components of the star's 
 motion (x and y in the plane of the sky), 
 it has the potential of determining all orbital parameters and the true mass of the planet.
 While astrometry on images is limited by the aperture size 
 and optical properties (distortions) of the telescopes used, astrometric measurements with 
 an interferometer have the potential to reach a much higher accuracy \cite{quir2001}.
\end{itemize}


\section{Astrometry with an interferometer} \label{sec-astrom}


\subsection{Wide-angle astrometry} \label{ssec-astrom-wa}

\begin{figure}
\begin{center}
\includegraphics[width=0.45\textwidth,angle=-90]{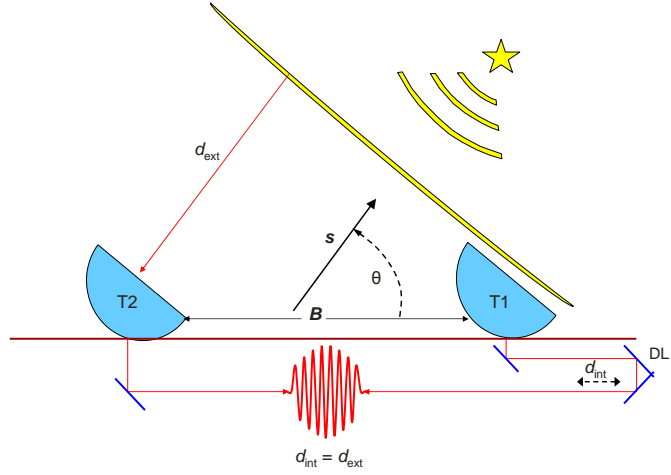}
\caption{\label{fig-astrom1}
  Single-star astrometry with an interferometer.
}
\end{center}
\end{figure}

A single-star interferometer measures the delay, $d$, 
(or OPD) between the sections of the wavefront from 
a star as they arrive at the two telescopes \cite{haniff2007,haniff2009}. 
This delay is given by:
\begin{equation}\label{eq-1}
d = \vec{s}\cdot\vec{B}+C = B\cdot\cos\theta+C \qquad ,
\end{equation}
where 
$\vec{s}$\ is the direction vector to the star,
$\vec{B}$\ is the baseline vector (and $B$\ its length), 
$\theta$\ is the angle between $\vec{s}$\ and $\vec{B}$, and 
$C$\ is a fixed instrumental term that must be calibrated (see Fig.\,\ref{fig-astrom1}). 
$\theta$\ corresponds to the position angle of the star on the sky, projected onto 
the baseline.
The external delay, which cannot be measured directly, is compensated with an 
optical delay line (DL) in the interferometer. At the position of the maximum 
stellar fringe, external and internal delays are equal. An approximation to the internal 
delay is measured with a laser metrology system.
Thus, by precisely knowing the interferometer baseline and measuring the delay, 
one can determine the position angle of the star with respect to the baseline 
with a precision that is no longer limited by the aperture size of the telescopes.
Atmospheric perturbations, however, introduce random piston fluctuations which add up 
to the the measured total delay. Although this could be partially overcome 
by observing at two wavelengths, it limits the astrometric accuracy to the mas level.


\subsection{Narrow-angle astrometry} \label{ssec-astrom-na}

\begin{figure}[htb]
\begin{center}
\includegraphics[width=0.48\textwidth,angle=-90]{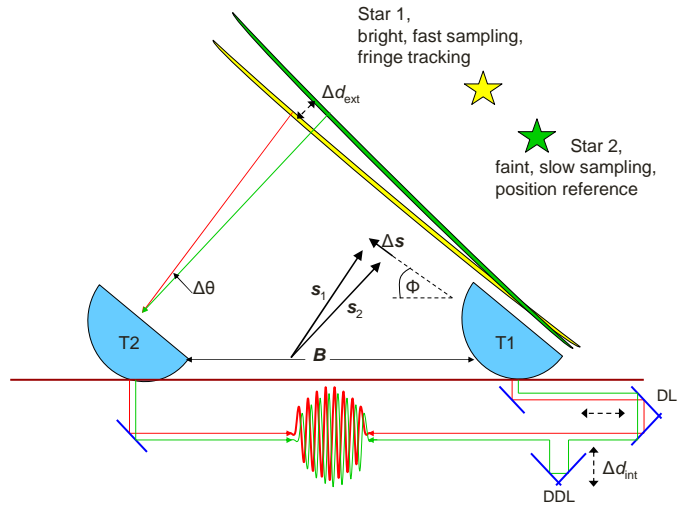}\\[2mm]
\caption{\label{fig-astrom2}
  Narrow-angle astrometry with a dual-star interferometer.
}
\end{center}
\end{figure}

To circumvent this problem, a dual-star interferometer like PRIMA measures only the 
differential delay between two stars, which corresponds to the angular separation 
between the two stars on the sky, projected onto the baseline. 
If the angular separation between the two stars is smaller than the isopistonic angle
of the atmosphere, where the piston perturbations and hence the fringe motions 
of the two wavefronts are correlated,  
the mean differential piston perturbations average to zero for sufficiently long 
observing time \cite{shao1992}. 
The differential delay, $\Delta d$, is then given by:
\begin{equation}\label{eq-2}
\Delta d = \Delta\vec{s}\cdot\vec{B}+\Delta C = \Delta s\cdot B \cdot \cos\phi+\Delta C \qquad, 
\end{equation}
where $\Delta\vec{s}$\ is the separation vector between the two stars in the plane of 
the sky, $\phi = 90^{\circ}-\theta$\ is the angle between the separation vector and 
the baseline vector, and $\theta$\ is the angle between the the baseline vector and 
the direction towards projected central position between the two stars 
(see Fig.\,\ref{fig-astrom2}).
The differential instrumental delay, $\Delta C$, is close to zero and can be calibrated.
In contrast to $\vec{s}$, $\Delta\vec{s}$\ is no longer a unit vector. 

The calibrated differential delay (Eq.~\ref{eq-2}) is then proportional to the 
projection of the separation vector between the two stars onto the interferometer
 baseline, i.e., it is a one-dimensional measurement.
In order to obtain both dimensions of the 
separation vector, one has to observe either with two different (orthogonal) baselines, 
or with one baseline at different parallactic angles.

This dual-star technique has another advantage besides providing a reference
against which positional offsets can be measured (narrow-angle astrometry).
If one of the two stars (the ``primary star'') is bright 
enough to measure its fringe phase within the atmospheric coherence time 
(fringe-tracking), it can be 
used to stabilize the fringes on the other (secondary) star (phase-referencing), 
thus allowing for much longer coherent integrations and hence increasing the 
limiting magnitude and the number of observable objects.


\subsection{Astrometry with PRIMA} \label{ssec-astrom-prima}

PRIMA will be the third optical\,/\,IR long-baseline dual-star interferometer 
that can perform narrow-angle astrometry as described above (see Sect. \ref{ssec-astrom-other}).
In its astrometry mode, PRIMA will mainly work with the 1.8\,m ATs, because they 
have much less severe vibration problems 
than the UTs. Phase-referenced imaging does not depend so much on the highest possible OPD 
accuracy and will, therefore, preferably be done with the larger UTs.

In PRIMA, Star Separator units (STS) located at the Coud{\'e} focus of each telescope 
separate two sub-fields (with a star in each) from the telescope field-of-view 
and send them as collimated beams through the interferometers. 
The two beams from each telescope are first sent parallel and close to each other 
through one main delay line (DL), 
to compensate the large delay that is common to both stars. 
This ensures that differential perturbations between the two beams introduced in the 
long DL tunnels are kept at a minimum. To obtain fringes from both stars on 
the detector, the differential OPD (dOPD) between the two stars must also be compensated. 
This is done with four Differential Delay Lines (DDLs; see Fig.\,\ref{fig-astrom2}), 
one for each star and telescope. 
Although the measurement principle would require only one DDL, operational reasons 
and symmetry requirements for the interferometer lead to the design with one DDL for each beam.
The dOPD for two stars separated by less than an arc\-min is much smaller 
(up to a few cm only) than the main OPD.
But, its compensation requires that the two beams travel different 
path lengths, giving rise to differential longitudinal dispersion (the metrology works at a 
different wavelength that the observed starlight) and other distortions that affect 
the dOPD measurements. Therefore, the DDLs operate in vacuum.

After the four beams have traveled through the main DLs and the DDLs, they arrive 
in the subterranean interferometric laboratory, where the Fringe Sensor Units (FSUs) 
and the other interferometric instruments are located (currently AMBER and MIDI). 
The beams from the bright star are combined and the fringe phase (for fringe-tracking) 
and group delay (for astrometry) measured in one of the two FSUs. 
This happens at a frequency of about 1\,kHz to sample the atmospheric coherence time of 
$\approx1$\,ms in K-band.
The FSUs combine the beams in the pupil plane, but they do not scan the fringes by 
modulating the OPD. Instead, the combined beam is split into four output beams 
that are shifted in phase with respect to each other 
(relative phases of the four beams: $0, \pi/2, \pi, 3\pi/2$).
The intensities of these four beams, from which the phase of the input beam is computed, 
are then measured by an infrared camera. 
The offset from zero phase of the bright star is used to control the DLs and DDLs and 
modify the internal OPD such 
that the phase is brought back to zero.
In astrometry mode, the beams from the other (fainter) star are combined in the same way 
in the second FSU. Since the OPD corrections, derived from the real-time fringe measurements 
on the bright star with the first FSU, are also applied to the beams of the faint star, 
this second FSU can now integrate much longer (several seconds) to detect the fringes 
without smearing them and measure their position. 
The total integration time of typically 30\,min is built up by continuously repeating this 
sequence.

A laser metrology system is used to measure the internal dOPD in the interferometer 
between the two stars. This H-band laser beam is injected into the star beams inside 
the FSU beam combiners, 
travels ``backwards'' through the entire VLTI optical train until it is reflected back at a 
retroreflector which is located between the STS and the telescope Coud{\'e} train with the derotator.
This retroreflector, or more precisely, its image in the entrance pupil of the telescope, is the reference 
point for OPD measurements which separate the differential delay $\Delta d$, 
measured by PRIMET, from the baseline, $\vec{B}$\ (Eq.\,\ref{eq-2}).

In order to be able to eliminate the constant term $\Delta C$\ in Eq.\,\ref{eq-2}
from the measurements, it is necessary to periodically exchange the two stars within 
the instrument during an astrometric measurement. This will be done by turning the 
field derotators in the AT Coud{\'e} trains by 180$^{\circ}$. 
This implies that calibrated delays cannot be derived 
before the observing sequence is finished.


\begin{figure}
\begin{center}
\includegraphics[width=0.5\textwidth]{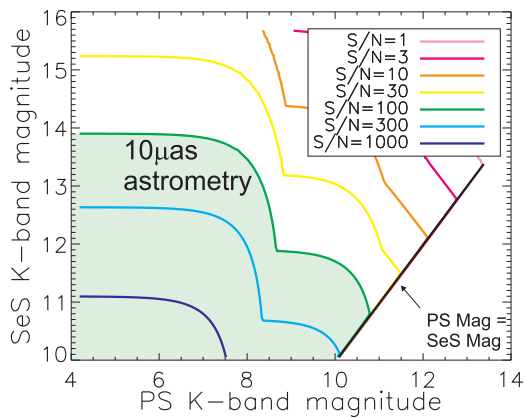}
\caption{\label{fig-ssmag}
  Dependence of astrometric accuracy of PRIMA with two ATs on brightness for
  10$^{\prime\prime}$\ separation between primary (PS) and secondary (SeS) star and
  30\,min integration \cite{tubbs2008,lau2008c}. Note that this figure is the 
  result of an error budget simulation, based on PRIMA subsystem specifications, 
  but without knowing the actual throughput and performance of all components. 
  These actual sensitivity of PRIMA must be verified on real stars during 
  commissioning in 2009.
}
\end{center}
\end{figure}


With two ATs, the minimum K-band brightness of primary stars, required for fringe-tracking 
is $K\approx 8$\,mag. 
The astrometric accuracy of PRIMA is directly related to the accuracy of the interferometric 
differential phase measurement. 
The S/N of the visibility phase is in turn proportional to the visibility amplitude of the 
interferometric fringes. A S/N of 100 of the detection of the cross-visibility would correspond 
to an interferometric differential phase error of 1/100 radians. At a wavelength of 2.2\,$\mu$m, 
this corresponds to an OPD error of 3.5\,nm. On a 150\,m baseline (typical for 
PRIMA astrometry), this corresponds to an astrometric error of $\approx 5\,\mu$as.
This accounts for about half of the anticipated OPD error of PRIMA, 
the other half coming mainly from seeing effects.
Therefore, a S/N of 100 for the fringe detection of the reference star has been assumed 
to be necessary for reaching an astrometric accuracy of 10\,$\mu$as.
In 30\,min integration time, this can be reached on a $K\le 14$\,mag star 
\cite{tubbs2008,lau2008c} (see Fig.\,\ref{fig-ssmag}). 
Note that both the limiting magnitude for fringe-tracking as well as the brightness constraint 
on reference stars are model estimates, which must be verified 
on real stars during the commissioning of the instrument in 2009.

The isopistonic angle at Paranal and in K-band is $\approx$10-20\,\arcsec; 
a value that will also be verified during PRIMA commissioning. 
The optical design of the STS does not allow to separate stars closer than 
2\arcsec\ without diverting light into the other channel. 
Therefore, there is not only a maximum angular separation, limited by the isopistonic 
angle of the atmosphere, but a minimum as well. 

For astrometry with PRIMA, AT baselines with 80 -- 160\,m length will typically be used.
On a 100\,m baseline, a separation of 10\arcsec\ between two stars corresponds 
to an OPD of $\approx$5\,mm in the interferometer, which must be compensated with the DDLs. 
The anticipated measurement accuracy 
of 10\,$\mu$as corresponds to a dOPD of only 5\,nm. 
This number defines the total dOPD error budget for PRIMA in the astrometric mode. 
Since the limitation due to residual atmospheric turbulence during a 30\,min integration 
will be of the same order, we expect to reach a final single astrometric measurement 
accuracy of $\approx 15-20$\,$\mu$as.


\subsection{Narrow-angle astrometry with other instruments} \label{ssec-astrom-other}

PRIMA is not the only instrument that can perform narrow-angle astrometry.
The first real narrow-angle astrometry measurements were done in 1992 
at the Mark\,III interferometer \cite{colavita1994}.
The PHASES project \cite{muter2005} at the Palomar Testbed Interferometer (PTI, \cite{boden2004}) 
has successfully demonstrated 100\,$\mu$as differential astrometry and two-combiner phase referencing.
It can perform differential astrometry on bright binary stars with separations in the range 
of 0.1-1.0 arcseconds.
The ASTRA upgrade of the Keck Interferometer will enable soon narrow-angle astrometry 
with an anticipated accuracy of $\approx 100\,\mu$as on star pairs with separations and 
brightness limits similar to PRIMA \cite{pott2009}.
GRAVITY, an interferometric imager for the VLTI with astrometric capability, 
but with a total field of view of only 1.7\arcsec, is currently being developed and will enable 
10\,$\mu$as very narrow-angle astrometry on galactic center stars after 2012 \cite{eisen2008}.
The most important upcoming astrometric space missions, Gaia and SIM\,Lite, 
will not rely on real narrow-angle (dual star) astrometry, but shall be mentioned here 
because in particular SIM also addresses similar scientific questions as PRIMA and my thus 
be relevant to extent the time baseline for certain PRIMA targets.


\section{Prospects for planet detections with astrometry} \label{sec-astrom-det}

The RV method is very efficient in detecting planets in 
short-period orbits close to the star. It requires stars with a sufficient number 
of narrow spectral lines,  i.e., fairly old stars (Gyrs) of about 1.2 M$_{\odot}$\ or less. 
More massive stars ($M > 1.4\,\msun$) as well as young and chromospherically active 
stars are often excluded from RV planet searches because their RV is more difficult 
to determine.
Stars $M > 1.4\,\msun$\ have only very few usable spectral lines.  
Young and active stars have often broad (rapid rotators) and unstable 
(chromospheric activity) lines.
Correspondingly, our knowledge on the RV planet population is biased towards planets 
in short and intermediate-period orbits around solar-type stars. 
RV measurements also provide no  
constraint on the inclination of the orbit ($sin\,i$), and thus usually 
only a lower limit on the planetary mass can be derived.
For this reason, astrometric orbit measurements are ultimately required to derive 
the fundamental parameter of a planet: it's mass. However, to play a significant 
role and open new discovery spaces, an astrometric accuracy of order 10--50\,$\mu$as 
is needed. 

The semi-amplitude of the RV variation, $K_1$, follows from Kepler's laws and 
is given by
\begin{equation}\label{eq-rv}
K_1 = \frac{m_p\,sin\,i}{(m_{\ast} + m_p)^{2/3}}\,\cdot\,\sqrt[3]{\frac{2\,\pi\,G}{P}}\,\cdot\,\frac{1}{\sqrt{1 - e^2}}\qquad,
\end{equation}
where $m_{p}$\ is the mass of the planet,
$m_{\ast}$\ is the mass of the star,
$i$\ is the inclination angle of the orbital plane against the line of sight,
$P$\ is the orbital period, 
$G$\ is the gravitational constant, and 
$e$\ is the orbital eccentricity.

Astrometry, on the other hand, is a complementary technique 
with a different detection bias. It favors planets in wide, 
long-period orbits (like in our own Solar System). Furthermore, astrometry can measure 
two components of the stellar reflex motion, versus the single radial component that 
is observable spectroscopically, thus allowing to derive full orbit solutions. 
From the definition of the barycenter of a two-body system, one can easily 
derive the semi-amplitude of the astrometric ``wobble'' of a star that is orbited by a planet:
\begin{equation}\label{eq-as1}
\rho  = \frac{m_p}{m_{\ast}+m_p}\,\cdot\,\frac{a}{D}\qquad,
\end{equation}
where $a$\ is the distance between the two bodies' centers and 
$D$\ is the distance towards the objects.
Invoking Kepler's third law and neglecting eccentricity (i.e., assuming $e=1$), gives:
\begin{equation}\label{eq-as3}
\rho = \frac{m_p\,(m_{\ast}+m_p)^{1/3}}{m_{\ast}}\,\cdot\,\sqrt[3]{\frac{P^2\,G}{4\,\pi^2}}\,\frac{1}{D}\qquad.
\end{equation}
%


\begin{figure}
\begin{center}
%
\includegraphics[width=0.9\textwidth]{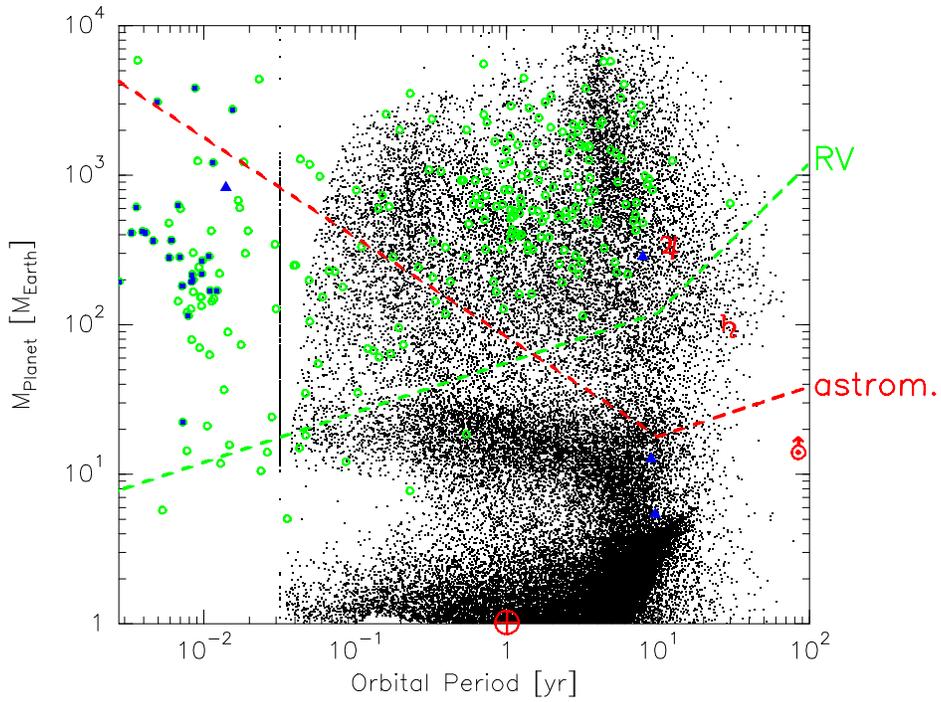}
\caption{\label{fig-plandet}
  Detection spaces (planet mass vs. orbital period) for planet searches with 
  astrometry and the RV technique. Black dots represent the results 
  of a planet population synthesis for solar mass stars, based on the core accretion paradigm 
  \cite{morda2008}. Inward migration was artificially stopped at 0.1\,AU, which 
  explains the pile-up line at 0.03\,yrs and the discrepancy between model and observations 
  for hot Jupiters.
  Known extrasolar planets 
  are marked as green circles (RV, lower mass limit due to $sin i$\ uncertainty), 
  blue squares (transit photometry), and blue triangles (microlensing). Solar system planets 
  are marked in red by their respective symbols. 
  Detection thresholds are marked by dashed lines under 
  the following assumptions: host star mass 1\,$\msun$, RV detection limit 5\,m/s, 
  astrometric detection limits 50\,$\mu$as at distance\,5 pc, maximum timeline of 
  observations 10\,yrs. Note that many of the known RV planets are detected around stars 
  with masses lower than 1\,$\msun$\ and therefore can appear below the RV detection 
  threshold for 1\,$\msun$\ stars.
}
\end{center}
\end{figure}


Equations\,\ref{eq-rv} and \ref{eq-as3} show the different detection biases 
of the two methods: 
RV ($K\propto P^{-1/3}$) is more sensitive to planets in short-period orbits, 
while astrometry ($\rho\propto P^{+2/3}/D$) favors planets in 
longer-period orbits. 
This is also demonstrated in Fig.\,\ref{fig-plandet}, which shows the detection limits 
for planet searches with astrometry and the radial velocity technique in a planet mass 
vs. orbital period diagram. 
To illustrate which method is sensitive to 
what kind of planets, the detection curves are overplotted on a planet population 
synthesis for solar mass stars, based on the core accretion paradigm \cite{morda2008}
as well as the already known extrasolar planets.
The diagram shows that narrow-angle astrometry with an accuracy of 50\,$\mu$as (5\,$\sigma$) 
opens a new discovery space only for orbital periods longer than 1--3\,yrs. 
Although the exact 
location of the crossing point between the two detection curves depends on the characteristics 
of the specific target 
stars (e.g., distance, mass, activity levels) and may shift over time (better spectrographs), 
this estimate has important implications for target selection and observing strategies 
of astrometric planet search programs. 

When equipped with Differential Delay Lines, PRIMA will be able to perform narrow-angle 
astrometry in K-band with a single-measurement accuracy of up to 10--20\,$\mu$as. 
Although this ultimate accuracy goal may only be reached after at least 1\,yr of data 
will have been used for long-term calibrations (see Sect.\,\ref{sec-espri}), 
it will be capable of detecting Saturn-mass planets around nearby main sequence stars of 
any spectral type, down to Uranus-mass planets in 1--5\,AU orbits around nearby M dwarfs. 
This peak of the ice giant distribution is outside the reach of current-day RV measurements.
Since astrometry does not depend on narrow and stable spectral lines, it will also 
be sensitive to Jupiter-like giant planets around young stars which are less suitable 
for the RV method.
Earth-like rocky planets are, however, still out of reach for current-day ground-based astrometry.


\section{The ESPRI project: astrometric exoplanet search with PRIMA} \label{sec-espri}

Starting in 2010, when PRIMA will be commissioned and operational, the ESPRI consortium 
(see Sect.\,\ref{sec-intro}) will use the facility with two ATs to carry out an 
astrometric {\bf E}xoplanet {\bf S}earch program with {\bf PRI}MA (``ESPRI''). 
ESPRI will address the following outstanding issues:
\begin{itemize}
\item Resolve the $sin\,i$\ uncertainty from planet masses found by RV
   surveys and derive accurate planet masses. This measurement is fundamental to 
   study the planetary mass function, in particular the upper mass cut-off where 
   the statistics is poor.
\item Confirmation of hints for long-period planets in RV surveys.
\item Measure the relative orbit inclinations in multiple planetary systems.
\item Inventory of planets around stars with different mass and age. Of particular 
   interest are the most nearby stars, irrespective of their spectral type, 
   as well as young stars with ages up to few hundred Myr.
\end{itemize}


\begin{table}[ht]
\caption[]{\label{tab-targets} Pre-selected ESPRI target stars}
\begin{tabular}[t]{llll}
\hline \hline \noalign{\smallskip}
 & RV: & CNS: & YS:  \\
\noalign{\smallskip} \hline \noalign{\smallskip}
Brightness$^{1)}$: & $K\le 8$\,mag       & $K\le 8$\,mag    & $K\le 8$\,mag  \\
Distance:          & $D\le 200$\,pc      & $D\le 15$\,pc    & $D\le 100$\,pc \\
Age:               & no restriction      & no restriction   & $a\,\le$\,300\,Myr  \\
Spectral types:    & F7 ... M3$^{2)}$    & A3...M6.5$^{2)}$ & B8...M3$^{2)}$   \\ 
Selection sources: & www.exoplanet.eu    & CNS$^{3)}$       & \cite{mamajek2002,montes2001,wich2003} \\
Number of stars$^{4)}$: &  148           & 367              & 380 \\
\noalign{\smallskip} \hline\noalign{\smallskip} 
\end{tabular}
{\footnotesize
\begin{list}{}{}
\item[$^{1)}$] Derived from the expected limit for hardware fringe-tracking. 
               Astrometry with reduced accuracy might be possible also on fainter stars 
               (see Fig.\,\ref{fig-ssmag}).
\item[$^{2)}$] No a priori restriction, but the limitations in brightness and distance constrain 
               the spectral types of the available stars. For group RV we also include 
               sub-giants (luminosity class IV).
\item[$^{3)}$] Unpublished version of the Catalogue of Nearby Stars \cite{jahr1993} 
               from August 2005 (Jahreiss, priv. comm.).
\item[$^{4)}$] Known spectroscopic binaries and visual binaries with projected separation $<$\,2\arcsec\ 
               are already excluded.
\end{list}
}
\end{table}


With these scientific goals in mind, three lists with potential 
target stars were preselected for the ESPRI project (see Table\,\ref{tab-targets}):
\begin{enumerate}
\item {\bf RV}: Stars with known RV planets within $\le$\,200\,pc 
                around the Sun.  
\item {\bf CNS}: Nearby stars of any spectral type within $\le$\,15\,pc around 
                 the Sun.
\item {\bf YS}: Young stars with ages $<$\,300\,Myr within $\le$\,100\,pc around the Sun.
\end{enumerate}

This preselected list contains in total about 900 stars. 
Of course, only stars that have suitable reference stars within the isopistonic angle 
can be observed with PRIMA. 
If such reference stars are available or not must be determined for every single 
star of this input list (see Sect.\,\ref{sec-prepobs}).
With the constraints on separation (2\arcsec\ -- $\approx$20\arcsec) and brightness 
($K\le 14$\,mag for SNR 100 in 30\,min integration time; 
see Sect.\,\ref{ssec-astrom-prima} and Fig.\,\ref{fig-ssmag}), one can expect to find 
on average one good target out of 10 candidates (10\%). 
Of course, the detection rate depends strongly on Galactic latitude, so the final ESPRI 
target list will be heavily biased towards the Galactic plane.
In total, the ESPRI project will monitor about 100 stars during about 200 nights spread 
over 5 years from 2010 through 2015.


\section{Necessary preparatory observations} \label{sec-prepobs}

Both narrow-angle astrometry as well as phase-referenced imaging of faint objects 
rely on the availability 
of reference stars within the isopistonic angle around the target object.
For faint source imaging this can be a real obstacle since the reference star must be 
bright enough for fringe-tracking. It is usually very unlikely to find such a bright 
star so nearby a pre-selected faint target object.

The situation is somewhat more favorable for narrow-angle astrometry of stars 
that are bright enough for fringe-tracking.
Although it strongly depends on the Galactic latitude, it is much more 
likely to find a fainter background star nearby any given foreground star than vice versa.
In this case, however, other problems become evident.

\begin{figure}[ht]
  \parbox{0.32\textwidth}{
  \begin{center}
  \epsfig{file={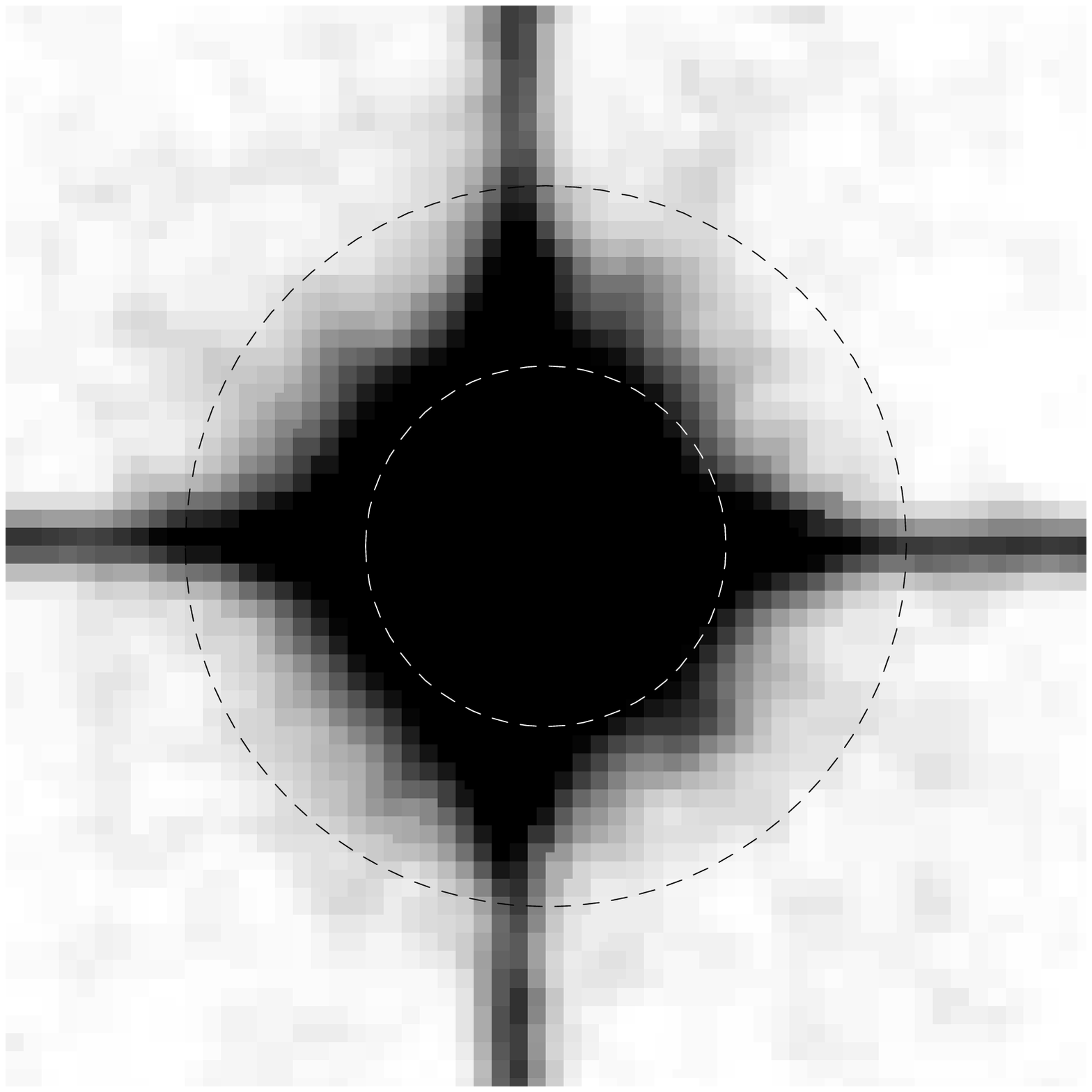},width=4.5cm}
  \end{center}}
  \hfil
  \parbox{0.32\textwidth}{
  \begin{center}\
  \epsfig{file={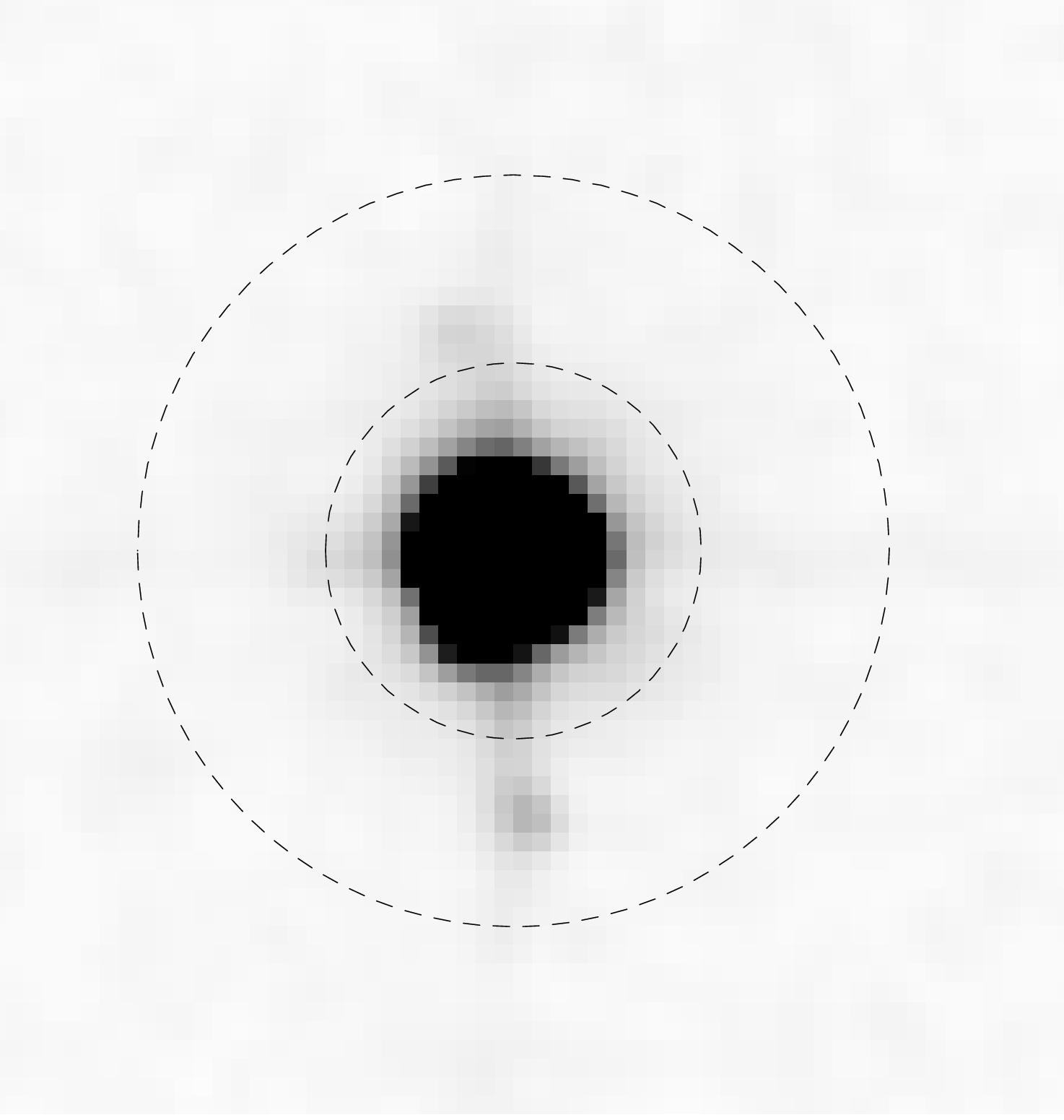}, width=4.5cm}
  \end{center}}
  \hfil
  \parbox{0.32\textwidth}{
  \begin{center}\
  \epsfig{file={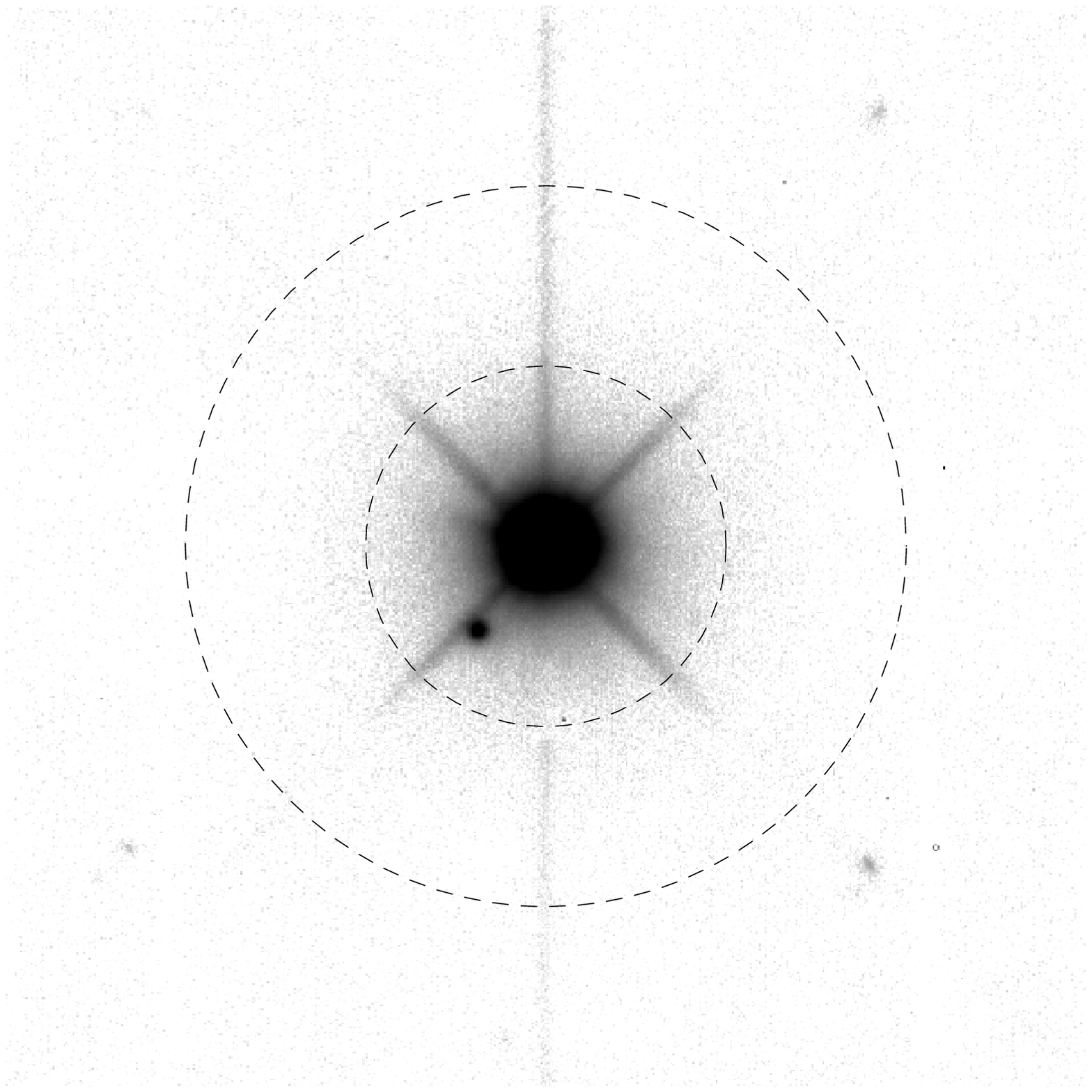}, width=4.5cm, angle=0}
  \end{center}}
 \caption[GJ745A]
  {\label{fig-sesim}
  Three images of HD\,87978, a potential ESPRI target star.
  Left: DSS2-red, Center: 2MASS-K, 
  Right: our high dynamic range K-band image obtained with SOFI at the ESO NTT in March 2008.
  North is up and East is left.
  The two circles denote radii of 10\arcsec\ and 20\arcsec\ around the target star.
  The star itself has $K\approx 6.5$\,mag. 
  The potential reference star at $r\approx6$\arcsec\ separation from the target   
  has $K\approx 12$\,mag. Objects such close to a very bright star are usually not detected 
  in 2MASS, DSS, or other all-sky surveys.
}
\end{figure}

Most public data bases are either not sensitive enough and therefore miss most of the 
potential reference stars, or they suffer from strong 
saturation and ``blind'' areas around bright stars ($K<8$\,mag or $V<3\ldots7$\,mag)
(e.g.,2MASS, Denis, DSS2, SDSS; see Fig.\,\ref{fig-sesim}). 
Others, like USNO B1.0 \cite{monet2003}, are 
so much dominated by ghosts and artefacts around bright stars that the false 
alarm rate is close to 100\%, while actually existing stars remain undetected. 
For these reasons it turns out that existing public data bases cannot be used 
to reliably identify reference stars for PRIMA observations.
Therefore, dedicated preparatory observations are a key pre-requisite for 
any PRIMA observation. 

Ideally, one would wish to use adaptive optics together with a coronographic near-infrared 
camera to hunt for faint background stars very close to very bright foreground stars.
However, such instruments are usually not available and would be ``too expensive'' 
for preparatory observations only. 
With dedicated ``standard'' NIR photometric imaging observations, optimized for 
high dynamic range, the problems of saturation effects and ghost images can be minimized 
(but not completely avoided) when the following advice is taken into account.
The total integration time should be built up by using the shortest possible single 
detector integration time (DIT), which is usually 1\,--\,2\,sec for many NIR cameras.
   This requires a camera with very fast readout scheme; otherwise the readout overhead will quickly 
   exceed the on-source integration time. Even then, the target stars will usually saturate, but 
   it should now be possible to identify faint background stars 
   as close as 1-3\arcsec\ to the bright star (see Fig.\,\ref{fig-sesim}).


An instrument that is currently (still) available at ESO and is suited to perform 
such a search for PRIMA reference stars is SOFI at the NTT at the La Silla observatory.
Figure\,\ref{fig-sesim} shows what can be done with SOFI when the above-mentioned rules 
are followed, as compared to all-sky surveys like DSS or 2MASS.

While the search for reference stars is a mandatory task when preparing any PRIMA observations, 
spectroscopic observations might be useful to characterize the target stars and to identify 
excessively active stars and spectroscopic binaries. 
The first ones could produce a too large astrometric ``noise''. 
Spectroscopic binaries are both in terms of fringe detection and interpretation and 
planet formation and orbit stability not well-suited to perform an astrometric planet search.


\section{Summary} \label{sec-sum}

This article gives an overview on how ground-based astrometric observations 
with an interferometer can be used to search for extrasolar planets. 
A facility that will soon become operational and is accessible to 
the ESO community is PRIMA. 

Sect. \ref{sec-planets} gives a general overview on planet search methods 
and, in particular, compares the two indirect methods 
RV technique and astrometry.

Sect. \ref{sec-astrom} explains how astrometric measurements are done with an interferometer, 
and how this is technically realized in PRIMA. 
They key points are:
{\it (i)} a dual-beam interferometer is needed to simultaneously observe 
  two stars that are located within the same isopistonic patch of the atmosphere, 
{\it (ii)} the interferometer measures the differential delay between the wavefronts from 
  the two stars, 
  and 
{\it (iii)} the differential delay and the interferometer baseline 
  are directly related to the part of the angular separation vector between the two stars that 
  is projected onto the baseline.
In order to reach the anticipated astrometric accuracy of 10-20\,$\mu$as, 
PRIMET metrology can measure the dOPD in the interferometer with an accuracy of 5\,nm.
The (narrow-angle) baseline will be determined with an accuracy of 
$\approx 50\,\mu$m. 
With two ATs, the minimum K-band brightness of primary stars, required for fringe-tracking, 
is $K\approx 8$\,mag. The minimum K-band brightness of reference stars required to reach 
10\,$\mu$as in about 30\,min integration time is $K\le 14$\,mag. 
The angular separation between the two stars must be 
\mbox{2\arcsec\,$< \Delta s < \approx$20\arcsec}.

Sect. \ref{sec-astrom-det} outlines the exoplanet detection space of 
a 10\,$\mu$as astrometric facility and compares it to the RV detection space.
Both methods have opposite detection biases with respect to the orbital period, 
with astrometry being more sensitive to longer period planets. 
In particular, narrow-angle astrometry with an accuracy of 50\,$\mu$as (5\,$\sigma$) 
opens a new discovery space only for orbital periods longer than 1--3\,yrs 
and can detect Uranus-mass planets in 1--5\,AU orbits around nearby M dwarfs. 
Earth-like rocky planets are, however, still out of reach for current-day ground-based astrometry. 

Sect. \ref{sec-espri} gives an overview on the astrometric planet search project that the 
ESPRI consortium wants to carry out with PRIMA between 2010 and 2015. Particular targets of 
the ESPRI programme are: 
{\it (i)} Stars with known exoplanets detected by RV,
{\it (ii)} the most nearby stars ($D\le 15$\,pc) around the Sun, and 
{\it (iii)} nearby young stars with $D\le 100$\,pc around the Sun.

Sect. \ref{sec-prepobs} describes which preparatory observations are necessary before 
a star can be astrometrically observed with PRIMA. In particular it is mandatory 
to search for reference stars around the potential target stars. Since available 
all-sky surveys often miss faint background stars close to very bright foreground stars 
(the typical planet search targets), dedicated high dynamic range NIR imaging observations 
are often unavoidable.





\end{document}